\documentclass[acus]{JAC2000}

\usepackage{graphicx}
\usepackage{epsfig}

\begin{document}
\title{HIGHER DIPOLE BANDS IN THE
NLC ACCELERATING STRUCTURE \thanks{Work supported by the US
Department of Energy contract DE-AC03-76SF00515.}}
\author{ C. Adolphsen, K.L.F. Bane, V.A. Dolgashev, K. Ko, Z. Li, R.
Miller, SLAC, Stanford, CA 94309, USA}

\maketitle

\begin{abstract}
We show that scattering matrix calculations for dipole
modes between 23-43~GHz  for the 206 cell detuned structure (DS)
are consistent with finite element calculations
and results of the uncoupled model.
In particular, the rms sum wake for these bands is comparable to
that of the first dipole band.
 We also show that for RDDS1 uncoupled wakefield calculations for
higher bands are consistent with measurements. In particular, a
clear 26~GHz signal in the short range wake is found in both results.
\end{abstract}

\section{INTRODUCTION}
In the Next Linear Collider (NLC)\cite{zdr},
 long trains of intense  bunches
are accelerated through the linacs on their way to the collision
point. One serious problem
that needs to be
addressed
is multi--bunch, beam break--up (BBU) caused by
wakefields in the linac accelerating structures. To counteract this instability the
structures are designed so that the dipole modes are detuned and
weakly damped.
Most of the effort in reducing wakefields, however, has been
focused on modes in the first dipole passband, which
overwhelmingly dominate. However, with a required reduction of
about two orders of magnitude, one wonders whether the higher
band wakes are sufficiently small.

For multi-cell accelerating structures higher band dipole modes
can be obtained by several different methods.
These include the so-called ``uncoupled'' model, which does not
accurately treat the cell-to-cell coupling of the
modes\cite{KBetal}, an open--mode, field expansion
method\cite{Yamamoto}, and a finite element method employing many
parallel processors\cite{Xiaowei}.
(Note that the circuit approaches\cite{Gluck}\cite{Jones} do not lend themselves well to
the study of higher bands.)
A scattering matrix  (S-matrix) approach can naturally be
applied to cavities composed of a series of waveguide
sections\cite{First}, such as a detuned structure (DS), and such a method
has been used before to obtain first band modes in detuned
 structures\cite{Ulla}\cite{Saml}. Such a method can also be
 applied to the study of higher band modes.

In this report we use an S-matrix computer
program\cite{Valery}\cite{pac99} to obtain modes of the 3rd to
the 8th passbands---ranging from 23-43~GHz---of a full 206--cell
NLC DS accelerating structure.
We then compare our results with those of a finite element
calculation and those of the uncoupled model. Next we repeat the
uncoupled calculation for the latest version of the NLC structure,
the rounded, detuned structure (RDDS1). Finally, we compare these results
with those of the DS structure and with recent wakefield
measurements performed at ASSET\cite{ASSET}.

%
\section{S-MATRIX WAKE CALCULATION}
 Let us consider an earlier version of the NLC accelerating
structure, the DS structure. It is a cylindrically--symmetric,
disk--loaded structure operating at X-band, at fundamental
frequency $f_0=11.424$~GHz. The structure consists of 206 cells, with
the frequencies in the first dipole passband detuned according to a
Gaussian distribution.
Dimensions of representative cells are given in
Table~\ref{str_par_tbl}, where $a$ is the iris radius, $b$ the
cavity radius, and $g$ the cavity gap.
Note that the structure operates at $2\pi/3$ phase advance,
and the period $p=8.75$~mm.

\begin{table}[htb]
\caption{Cell dimensions in the DS structure.}
\begin{center}
\begin{tabular}{|c|c|c|c|} \hline
 cell\# & a [cm] & b [cm]   & g [cm] \\ \hline
  1 &  .5900& 1.1486&  .749 \\
  51 & .5214 &1.1070 & .709 \\
103 & .4924 &1.0927 & .689 \\
154 &.4660 &1.0814 & .670 \\
206 &.4139 &1.0625 &.629 \\\hline
\end{tabular}
\label{str_par_tbl}
\end{center}
\end{table}

For our S-matrix calculation we follow the approach of
Ref.~\cite{pac99}: A structure with $M$ cells is modeled by a set
of $2M$ joined waveguide sections of radii $a_m$ or $b_m$, each filled
with a number of dipole TE and TM waveguide modes.
First the S-matrix for the individual sections is obtained, and
then, by cascading, the S-matrix for the
composite structure is found.
Using this matrix, the real part of
 the transverse impedance $R_\perp$ at discrete frequency points
 is obtained.
We simulate a structure closed at both ends, and one with no wall
losses. For such a structure $R_\perp$ consists of a series of
infinitesimally narrow spikes. To facilitate calculation
 we artificially widen them by introducing a small
imaginary frequency shift, one small compared to the minimum
spacing of the modes. To facilitate comparison with the results of
other calculation methods, we fit $R_\perp(\omega)$ to a sum of
Lorentzian distributions, from which we extract the mode
frequencies $f_n$ and kick factors $k_n$. Knowing these the wakefield
is given by
\begin{equation}
W_\perp(s)= 2\sum_{n} k_n \sin (2\pi f_n s/c)\,e^{-\pi f_n
s/Q_n c}\quad, \label{wake}
\end{equation}
with $s$ the distance between driving and test particles, $c$ the
speed of light, and $Q_n$ the quality factor of mode $n$.

\begin{figure}[htb]
\centering
\includegraphics*[width=72.5mm]{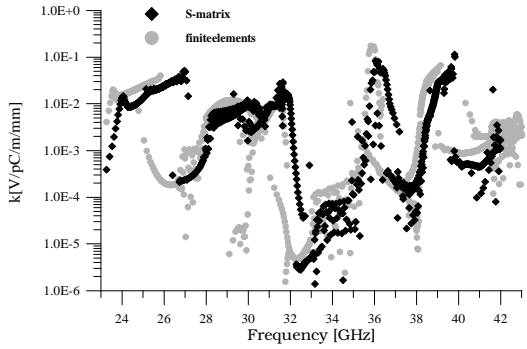}
\caption{Results for the DS structure as obtained by the S-matrix
and the finite element approaches. Note that the dimensions in the
two cases differ slightly.
} \label{kick_s_x}
\end{figure}

For our DS S-matrix calculation we approximate the rounded irises by
squared ones. We use 15 TE and 15 TM waveguide modes for each structure cavity
region, and 8 TE and 8 TM modes for each iris region. Our
imaginary frequency shift is 1.5~MHz. Our resulting kick factors, for frequencies
in the 3rd--8th passbands (23--43~GHz), are shown in
Fig.~\ref{kick_s_x}. (Note that the effect of the 2nd band modes is small and
can be neglected.)
In Fig.~\ref{kick_s_x} we show also, for comparison, the results
of a finite element calculation of the entire DS
structure\cite{Xiaowei},
an earlier calculation that, however, {\it does} include the
rounding of the irises.

We note from Fig.~\ref{kick_s_x}
that the agreement
in the results of the two methods is quite good, taking into account the difference in
geometries.
We see that the strongest modes are ones in the 3rd band
(24-27~GHz), the 6th band (35-37~GHz), and the 7th band
(38-40~GHz), with peak values of $k=.04$, .08, and .08~V/pC/mm/m,
respectively (which should be compared to .4~V/pC/mm/m for 1st
band modes).
However, thanks to the variation in $a$ (for the 7th
band) and $g$ (for the 3rd and 6th bands), these bands are seen to be
significantly detuned, or spread in frequency.
Another comparison is to take $2\sum k_n$ for the bands, a
quantity that is related to the strength of the wakefield for
$s\sim0$,
before coupling or detuning have any effect.
For our S-matrix calculation for bands 3-8 this sum equals 19, for the finite
element results 21~V/pC/mm/m (for the first two bands it is
74~V/pC/mm/m).

It is also necessary to know the mode $Q$'s to know the strength
of the wakefield at bunch positions. A pessimistic estimate takes
the natural $Q$'s due to Ohmic losses in the walls for the closed
structure. Assuming copper walls these $Q$'s are very high
 for some of these higher band modes ($>10000$). In the real
 structure, however, the $Q$'s can be much less, depending on the coupling of the
 modes to the beam tubes and the fundamental mode couplers,
 effects that in principle can be included in the S-matrix
 calculation. In practice, however, these calculations are very
 difficult.



\section{THE UNCOUPLED MODEL}
The uncoupled model is a relatively simple way of estimating the
impedance and the wake. It can be applied easily to higher band
modes (unlike the circuit models) and to structures that are not
composed of a series of waveguide sections (unlike the S-matrix
approach).
However, since it does not accurately treat the
cell-to-cell coupling of the modes, it does not give the correct
long time behavior of the wakefield.

The wakefield, according to the uncoupled model, is given by
an equation like Eq.~\ref{wake}, except that the sum is over the
number of cells $M$ times the number of bands $P$,
and the mode frequencies and kick factors are replaced by
${\tilde f}_{pm}$ and ${\tilde k}_{pm}$, which represent the synchronous mode frequencies
and kick factors, for band $p$, of the periodic structure with dimensions of cell
$m$ of the real structure. For
our uncoupled calculation we obtain the ${\tilde f}_{pm}$ and ${\tilde k}_{pm}$ for
a few representative cells of the structure using an electromagnetic field
solving program, such as MAFIA\cite{mafia}, and obtain them for
the rest by interpolation.

\begin{figure}[htb]
\centering
\includegraphics*[width=72.5mm]{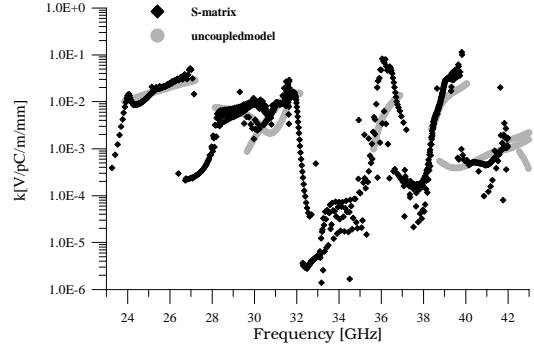}
\caption{ Kick factor comparison for the DS structure (square irises).
} \label{kick_s_u} 
\end{figure}

\begin{figure}[htb]
\centering
\includegraphics*[width=72.5mm]{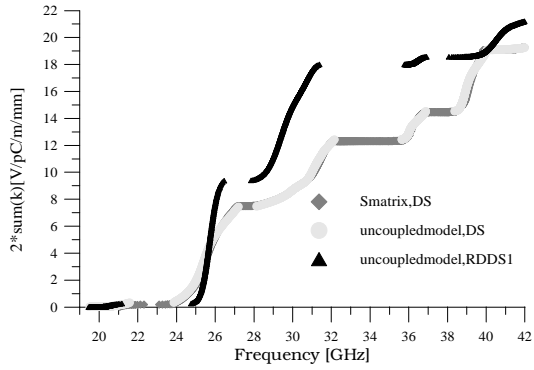}
\caption{Running sum of kick factor comparison.
} \label{sum_k_sk_bw} \vspace*{-3mm}
\end{figure}

In Fig.~\ref{kick_s_u} we plot again the kick factors obtained by
the S-matrix approach for the DS structure (rectangular irises),
but now compared to the results of the uncoupled model applied to
the same structure. The agreement is better than in
Fig.~\ref{kick_s_x}. We expect the kick factors for the two methods
 to be somewhat different, due to the cell-to-cell coupling, but the running sum of
kick factors, which is related to the short-time
wake, should be nearly the same. The running sum, beginning
at 20~GHz, of the two calculations is plotted in
Fig.~\ref{sum_k_sk_bw}. We note that agreement, indeed, is
very good.

\begin{figure}[htb]
\centering
\includegraphics*[width=72.5mm]{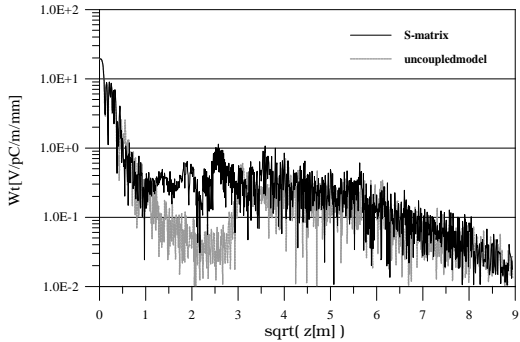}
\caption{Comparison of wakefields (23-43 GHz frequency range
only) for the DS structure (square irises).
} \label{wake_sk}\vspace*{-3mm}
\end{figure}

In Fig.~\ref{wake_sk} we plot the amplitude of the dipole wakes,
for the frequency range 23-43~GHz only, of the DS structure (with squared irises),
as obtained by the two approaches. (Here $Q$ has been set to 6500,
appropriate for copper wall losses for the 15~GHz passband). Note
that horizontal axis of the graph is $\sqrt {s}$ in order to
emphasize the wake over the shorter distances. Far right
in the plot is equivalent to $s=80$~m, the NLC bunch train length.
We note that the initial drop-off and the long-range wake are very
similar, though there is some difference in the region of 1-10~m.
The amplitude at the origin, 20~V/pC/mm/m, is small compared to
78~V/pC/mm/m for the first dipole band, but the longer time
typical amplitude of $\sim1$~V/pC/mm/m is comparable to that of
the first band. The rms of the sum wake, $S_{rms}$, an indicator of the
strength of the wake force at the bunch positions, for the higher bands is
.5~V/pC/mm/m, which is comparable to that of the first dipole
band. Depending on the external $Q$ for the structure, however, $S_{rms}$
for the higher bands may in reality be much smaller.


The latest version of the NLC structure is RDDS1 which has
rounded irises as well as rounded cavities. As such it is
difficult to calculate using the S-matrix approach. We have not
yet done a parallel processor, finite element
calculation, but we have done an uncoupled one. The sum
of the kick factors of the result is given also in
Fig.~\ref{sum_k_sk_bw} above. Although the running sums at 42~GHz
for DS and RDDS1 are very similar, at lower frequencies
the curves are quite different. In particular, the 3rd band modes
($\sim 26$~GHz) appear to be less detuned for RDDS1, the 4th
and 5th
band modes (27-31~GHz) are stronger, though still detuned, and
between 32-40~GHz there is very little impedance.

\section{ASSET MEASUREMENTS}
\begin{figure}[htb]
\centering
\includegraphics*[width=72.5mm]{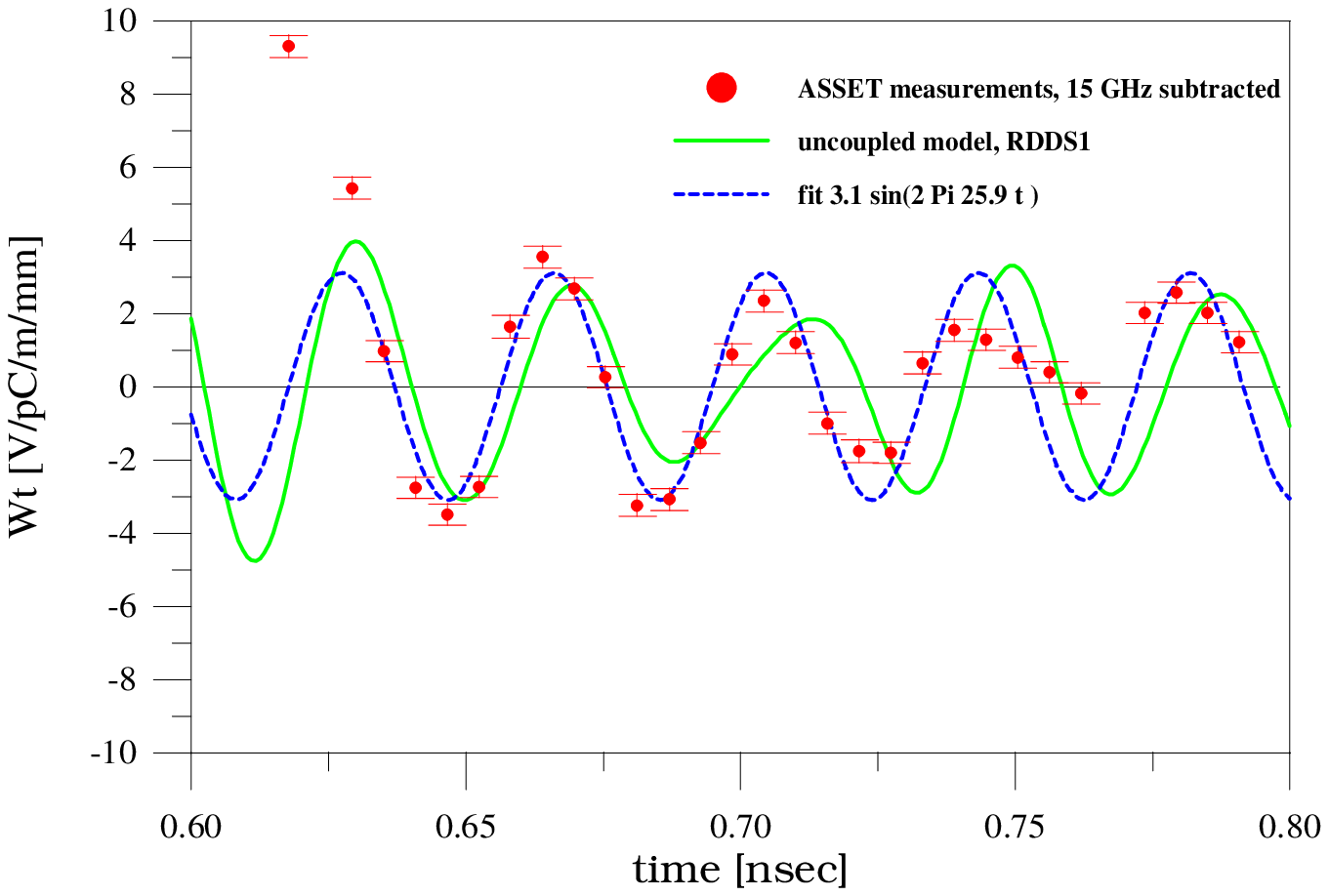}
\caption{The measured wake function for RDDS1, with the 15~GHz
component removed.
}
 \label{wk_ft07}
\end{figure}

\begin{figure}[htb]
\centering
\includegraphics*[width=72.5mm]{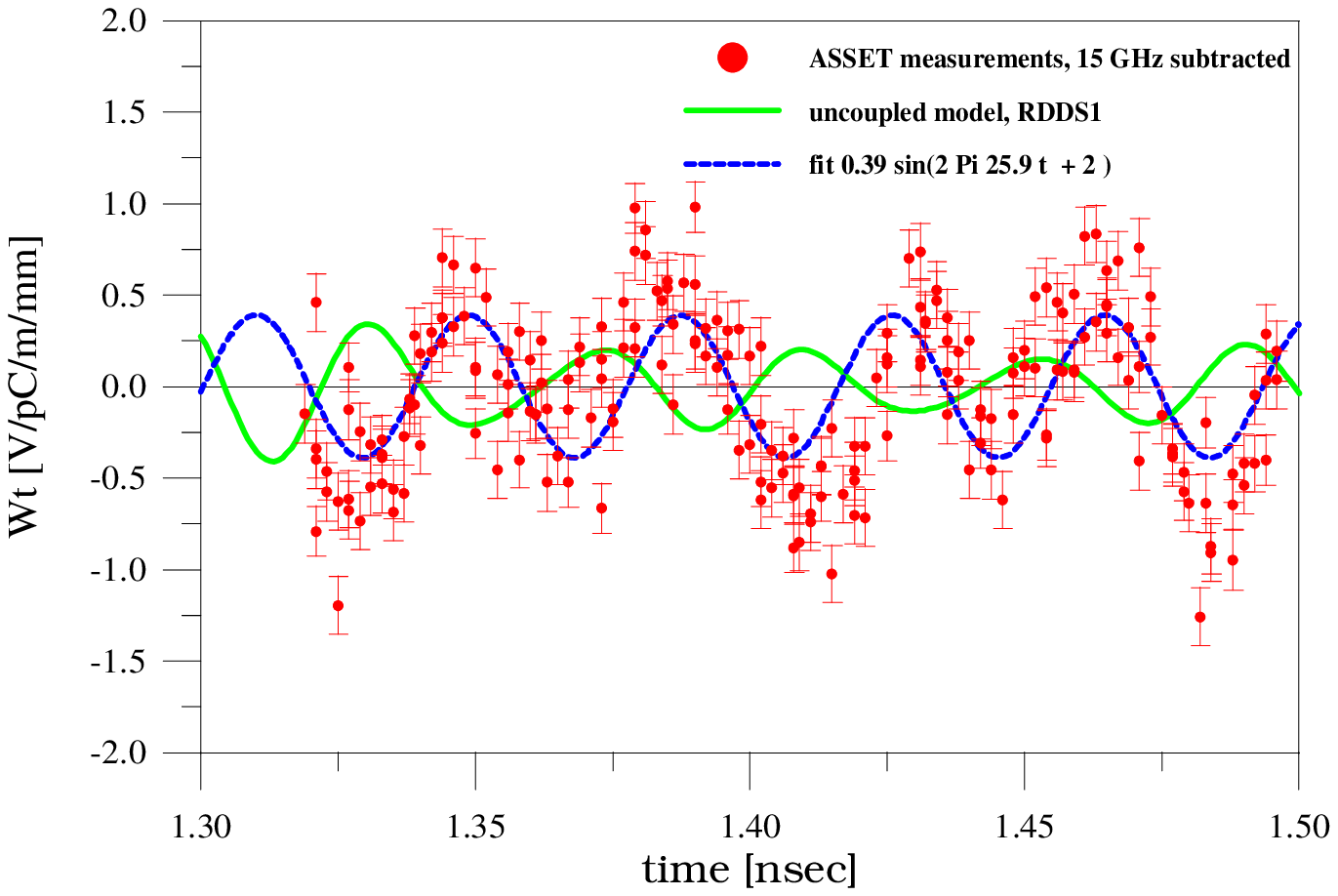}
\caption{The measured wake function for RDDS1, with the 15 GHz component
removed.
}  \label{wk_ft14} 
\end{figure}

Measurements of the wakefields in RDDS1 were performed
at ASSET\cite{ASSET}.
In Figs.~\ref{wk_ft07},\ref{wk_ft14} we
 present results for the vicinity of .7 and 1.4~nsec behind the
driving bunch.
To study the higher band wakes we have removed the 15~GHz
component from the data in the plots. The remaining wake was fit
to the function $A\sin(2\pi F+\Phi)$ with $A$, $F$, and $\Phi$
fitting parameters. This fit, along with the 3rd band component
 of the uncoupled
model results ($\sim 26$~GHz), are also given in the figures. At .7~nsec
this component is clearly seen in the data, and the amplitude and
phase are in reasonable agreement with the calculation.
At 1.4~ns there is more noise, though the 26~GHz component can
still be seen.

\end{document}